\documentclass[aps,prl,superscriptaddress,showpacs,twocolumn,floatfix]{revtex4}
\usepackage[dvips]{epsfig}
\usepackage{draftcopy}
\usepackage{color}
\begin{document}

\title{Conformational dependence of a protein kinase phosphate transfer
reaction}

\author{Graeme Henkelman}
\author{Montiago X. LaBute}
\author{Chang-Shung Tung}
\author{P.\ W.\ Fenimore}
\author{Benjamin H. McMahon}
\affiliation{Theoretical Division, Los Alamos National Laboratory, Los Alamos, New Mexico 87545} 
\date{\today}

\begin{abstract}

Atomic motions and energetics for a phosphate transfer reaction
catalyzed by the cAMP-dependent protein kinase (PKA) are calculated by
plane-wave density functional theory, starting from structures of
proteins crystallized in both the reactant conformation (RC) and the
transition-state conformation (TC).  In the TC, we calculate that the
reactants and products are nearly isoenergetic with a 0.2 eV
barrier; while phosphate transfer is unfavorable by over 1.2 eV in the
RC, with an even higher barrier.  With the protein in the TC, the
motions involved in reaction are small, with only P$_\gamma$
and the catalytic proton moving more than 0.5~\AA{}.
Examination of the structures reveals that in the RC the active site
cleft is not completely closed and there is insufficient space for
the phosphorylated serine residue in the product state.  Together,
these observations imply that the phosphate transfer reaction occurs
rapidly and reversibly in a particular conformation of the protein, 
and that the reaction can be gated by changes of a few tenths of
an \AA\ in the catalytic site.

\end{abstract}

\maketitle

\section {INTRODUCTION}

Protein kinases regulate many biological processes by transferring a
phosphate group from adenosine triphosphate (ATP) to the sidechains of
particular serine, threonine, or tyrosine residues.  The bulky,
charged phosphate group alters the conformation and function of the
target protein \cite{smith92,johnson01}.  Different kinases recognize
different primary sequence motifs surrounding the residue to be
phosphorylated, in a highly regulated fashion \cite{pearson91,
songyang94, pinna96, brink03}.  Structural studies have revealed
several conformational changes, such as closing of the active-site
cleft, the packing of the activation loop, and rotation of the
C-helix, which are often implicated in controlling the activity of
protein kinases \cite{johnson01}.  The reasons for such control are
clear, but no answer has been provided to such questions as: ``How
closed is closed?'', or ``is this particular conformation of the
activation loop `good enough' for phosphorylation to occur?''  Quantum
chemistry is required to objectively answer these questions.

The extent of conformational heterogeneity in a covalent protein
reaction was first quantified in a series of experiments monitoring
the temperature-dependent rebinding of CO to myoglobin after flash
photolysis \cite{fra75}.  Agmon and Hopfield created a concise
phenomenological model describing this situation, using transition
state theory to describe the vibrational reaction, and a diffusive coordinate which
describes the protein conformation and modulates the reaction barrier to the vibrational
transition \cite{agmon83a,agmon83b}.  Moving beyond the
phenomenological model requires specifying both the conformational
heterogeneity and the sensitivity of the reaction barrier to this
heterogeneity.  Careful structural analysis \cite{voj99} and quantum
chemistry calculations \cite{mcm00} showed that the reaction barrier
heterogeniety is indeed a reasonable consequence of observed
structural heterogeniety at the myoglobin active site.  A reevaluation
of a wide variety of myoglobin data also shows that a distinction
between diffusive, solvent-controlled conformational motions and
Arrhenius transitions which are independent of the solvent dynamics is
well supported by experiments \cite{fenimore02}.

In this work, we explore the conformational sensitivity of the protein
kinase reaction in two experimentally determined structures of PKA,
one crystallized with ATP and the protein kinase inhibitor (PKI),
which we refer to as the reactant conformation of PKA, or RC
\cite{zheng93}.  The other conformation is obtained by crystalizing
with a transition-state analogue, a non-reactive ADP-AlF$_3$ and a
mimic of PKI which is both shorter and has a phosphate-accepting
serine instead of an inert alanine at the reactive position
\cite{madhusudan02}.  We refer to the protein conformation in this
case as the transition state conformation of PKA, or TC. Although
differences in conformation between the two structures are $\sim$ 0.5
\AA, we find qualitatively different energetics of reaction, which
lead us to conclude that conformational motion of the protein kinase
is rate-limiting to the overall phosphate transfer reaction.

\section {METHODS}

Initial equilibrium geometries are generated from the protein data
bank entries 1ATP, (RC) and 1L3R, (TC), both crystal structures of PKA. Initial
guesses for the complete reactant and 
product structures were obtained by homology-modeling the terminal
phosphate of ATP (P$_{\gamma}$) and side chains of the serine and
catalytic aspartic acid (D-166) into appropriate positions.  All atoms
which were modeled in this way were allowed to move in subsequent
geometry optimization steps.  Four
different model sizes, containing 82, 88, 244 and 263 atoms, were
constructed.  The 82-atom minimal system has been defined by Valiev
{\it et al.} \cite{valiev03}, and includes the Mg$_2$-tri phosphate
and its immediate interaction partners, G-52, S-53, K-72, D-166,
K-168, N-171, D-184, and the substrate serine residue.  Two water
molecules from the crystal structure are included to complete the
six-fold coordination of Mg in the 88-atom system.  The 244-atom
system contains a significantly larger shell around the reaction
center and more of the ATP molecule (including part or all of residues
G-52, S-53, F-54, K-72, D-166, L-167, K-168, P-169, E-170, N-171,
D-184, F-185, G-186, and the substrate backbone from residue 18 to
22).  The atomic coordinates of backbone carbon and nitrogen atoms for
each system were taken directly from the experimental crystal
structures. Non-backbone bond lengths and most angles are derived from amino acid
templates to facilitate comparison of energies between
structures. Protons were added where needed. Residues were truncated
by replacing carbon atoms with protons and adjusting the new C-H bond
length accordingly. Full atomic coordinates are provided in
the supporting material.

Atomic interactions are described by the VASP density functional theory (DFT) 
and a plane waves basis \cite{kresse93,kresse96}.  Ultrasoft
pseudopotentials of the Vanderbilt form \cite{vanderbilt90}, and a
PW91 generalized gradient approximation functional are used.  A 270 eV
plane wave cutoff, appropriate for the pseudopotentials, is applied.
This cutoff was increased to 300 eV to verify insensitivity of results to
choice of basis set.  The reaction pathways are computed with periodic images separated by
at least an 8 \AA\ vacuum layer. A periodic box size of
$19\times21\times16$ \AA$^3$ is used for the 82- and 88-atom clusters,
$20\times24\times21$ \AA$^3$ for the 244-atom cluster and $27\times
24\times 23$ \AA$^3$ for the 263 atoms structure. 

Reactant and product geometries are calculated by optimizing the geometry of 
the clusters with a conjugate gradients algorithm.  Saddle points connecting 
these stable geometries were found using the nudged elastic band (NEB) method 
\cite{jonsson98}.  In this method, images are generated by linear interpolation
between the optimized reactant and product structures, and DFT is used to 
calculate forces on each atom of each image.  The images are then geometry
optimized subject to harmonic forces between the images which force them to 
be equally spaced along a 
minimum energy pathway between reactants and products.  The climbing
image modification to the NEB method \cite{henkelman00,henkelman00a}
was used so that the highest energy image along the band converges
directly to the saddle point, thus increasing the accuracy of the energy
barrier with \textcolor{red}{fewer images}.  Refinements to the
barriers for small 
system changes were computed with the dimer saddle point finding
method \cite{henkelman99}.

The 82-atom system contains the same 59 unconstrained atoms as Valiev
{\it et al.}, while the waters added to make the 88-atom system were allowed
to move.   The 244-atom system has only 42 moving atoms (the
gamma phosphate, Mg1, Mg2, their coordinated water, and the sidechains
of S-21, D-166, and K-168).

In order to test the sensitivity of the 244-atom result to system
size changes and details of the electrostatic boundary conditions and
periodic box size, we added the adenosine ring to make a 263 atom system, and
observed changes to the barrier for TC of less than 15 meV.

\section {RESULTS}

\subsection{conformational dependence}
\begin{figure}[t!]
\center
\includegraphics[width=8.5 cm]{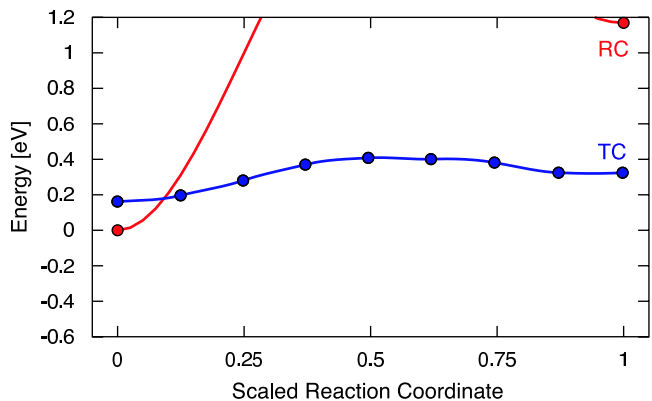} 
\vskip 0.2cm
\includegraphics[width=8.0 cm]{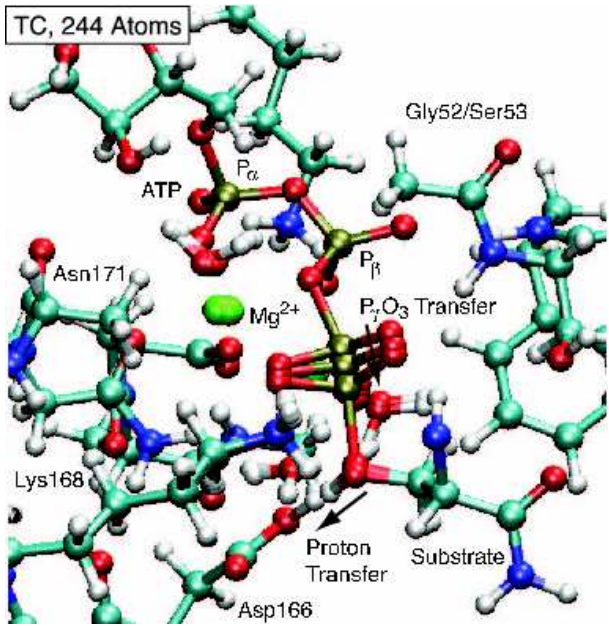}
\caption{(a) Energy barrier for the 244-atom system in the RC and TC
conformers.  With the additional constraining atoms, the RC reaction
becomes very unfavorable, while the TC conformation
phosphorylates. (b) Reaction path for phosphate transfer in PKA for a
244-atom TC structure.  The additional constraining atoms on the
outside of the cluster, as compared to the 82-atom system in
Fig.~\ref{pka82_fig}(a), reduce spurious motion during the reaction.
For example, the central Mg$^{2+}$ ion remains in place with the
proper coordination throughout this reaction.
}
\label{244atom}
\end{figure}

Fig.\ \ref{244atom}a shows the energetics of the 244-atom systems.
The reaction is endothermic in RC by 1.2 eV, while in TC, the reaction
is nearly isoenergetic with a barrier of only 0.20 eV.  The reaction
pathway of TC, shown in Fig.\ \ref{244atom}b, shows only small motions
during the reaction; only two atoms move more than 0.5 \AA, and the
catalytic base moves only 0.07~\AA.  Apparently, in the TC structure,
the atoms around the active site are correctly arranged for both the
reactants and products state. In contrast, 1.2 eV endothermicity in RC
indicates sub-optimal geometries in the product state.

At the transition state, the P$_\gamma$O$_3$ has a planar
configuration, halfway between the ADP and the serine O$_\gamma$, and
proton transfer from the serine to the Asp-166 occurs after the
phosphate transfer. 
Transferring the proton to an oxygen atom on the P$_\gamma$ phosphate
was found to be unfavorable by more than 0.5 eV.  These observations are in
agreement with previous DFT studies \cite{valiev03}.

One possibile spurious origin for the low barrier in TC is that both 
the reactants and products are destabilized because the active
site has collapsed around the AlF$_3$, just as we expect RC to be energetically
biased towards the reactants.  Two observations force us to the conclusion
that this is not occuring.

Examination of the reaction pathway geometries reveals two clear reasons
for the endothermicity in the TC structure.  First, the reaction cleft in RC is
approximately 1~\AA\ more open in RC than in TC, so the protein is
unable to maintain the full octahedral coordination of both magnesium ions in
the products state.  Fig.~\ref{pka82mg_fig} shows this distance
between the N-171 oxygen and the closest P$_\beta$ oxygen to
be 5.1~\AA\ in the RC conformer.  In the TC structure, the reaction
center is compressed, reducing this distance to 4.2~\AA.  Second,
it appears that there is more space for the phosphorylated serine residue
in TC than RC.

Fig.\ \ref{244atom}a maintains the relative energies of RC and TC,
and shows that TC is only 160~meV higher than RC.  The errors
contributing to this difference are likely several tenths of an eV,
but it is gratifying that the reactants energies are so similar in the
two conformations.

\subsection{geometry constraints}
Another question arises from the relatively small set of moving atoms
allowed in the calculation shown in Fig.\ \ref{244atom}.  One might
expect the geometry-optimized reaction pathway to be independent of
initial conformation in the limit of the entire protein being allowed
to move.  Several observations force us to the conclusion that this is
not correct.

\begin{figure}[t!]
\center
\includegraphics[width=8.5 cm]{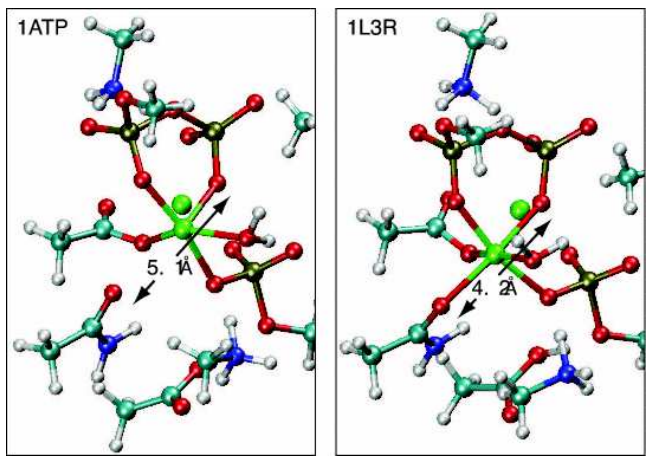} 
\caption{The 82-atom system in the TC structure has a lower barrier
than the RC structure and a favorable product state because the
reaction center is compressed.  In the product and transition states,
the central Mg atom can remain octahedrally coordinated in the TC
conformer.  In the RC conformer the reaction cleft is opened and the
same Mg atom is forced to break one bond, making the reaction
unfavorable.  The distance between the coordinating oxygen atoms on
the Asn171 and ADP groups, which is a measure of reaction center size,
is increased from 4.3~\AA\ in the TC structure to 5.0~\AA\ in the
RC structure.}
\label{pka82mg_fig} 
\end{figure}

First, energetic minimization of both RC and TC towards the reactants
state, using a classical molecular dynamics potential (which should be
perfectly valid for the reactants state) causes only one third of the
difference in the distance shown in Fig.\ \ref{pka82mg_fig} to
dissappear, even with no solvent present and every atom in the protein
allowed to move during the optimization.

Second, Table~I, shows much smaller motions are needed for reaction in
TC than in RC.  The Mg$^{++}$ ions, coordinated water molecules,
catalytic base, and lysine all move $\sim$ 0.1~\AA\ in TC, and 0.5 to
1.0~\AA\ in RC.  Even the serine residue, which gains a phosphate
group, moves only a quater \AA\ during the course of reaction.  A
related observation is that the change in force on the frozen atoms
during reaction is about half as large for TC than RC (not shown).

\begin{table}[t!]
\begin{center}
\begin{tabular*}{8.5cm}{l|r|r|r|r}
\hline \hline
 & 82 RC & 244 RC & 82 TC & 244 TC \\
\hline
Mg1 & 0.63  & 0.16 & 0.61  & 0.10 (0.04) \\
Mg2 & 1.07  & 1.04 & 0.42  & 0.25 (0.16) \\
\hline
P$_{\gamma}$ & 1.65  & 1.10 & 1.20 & 0.89 (0.56) \\
O$_{1\gamma}$ & 1.20 & 0.53 & 0.72 & 0.35 (0.22) \\
O$_{2\gamma}$ & 1.43 & 0.67 & 0.64 & 0.33 (0.19) \\
O$_{3\gamma}$ & 0.99 & 0.42 & 0.56 & 0.49 (0.37) \\
\hline
Ser O$_{\gamma}$ & 1.43 & 0.43 & 1.52 & 0.27 (0.25) \\
H & 1.24 & 1.35 & 1.25 & 0.53 (0.07) \\
Asp O & 1.12 & 0.57 & 0.96 & 0.07 (0.04)  \\
\hline\hline
\end{tabular*} 
\end{center}
\caption{\label{move_tab}The distances in \AA\ that particular atoms
moved in the four reaction pathways. The distance from the reactants
minimum to the transition state is indicated in parenthesis for the
244-atom TC structure.}
\end{table}

To directly check the effect of constraint on the results, and guided
by knowlege of the forces on frozen atoms, we relieved the constraint
on the beta phosphate of ATP.  The relative energy of RC and TC
went up by 350 meV, while the maximum change
in force dropped by a factor of ten, and was uniformily distributed across
a dozen boundary atoms.  The exothermicities of RC and TC changed by less than
0.1~eV, while optimal geometries differed by $\sim$ 0.1~\AA, suggesting
that the templates used to build the Mg-triphosphate coordination
sphere differed from the quantum mechanical potential by $\sim$ 0.1
\AA\ in several places.

The conclusion that TC provides a structure with happy reactants as
well as the potential to create happy products with minimal motions is
robust.

\subsection{dependence on system size}
Valiev et al. \cite{valiev03} reported a barrier of 0.5 eV for this reaction,
when using either a GGA or B3LYP functional and local basis set on an 82-atom
version of the RC structure, with only a pair of atoms fixed at the boundary
of each molecular fragment.  This result is intermediate between our
RC and TC results, and could indicate computational details provide variability
as large as the conformational differences which we cite.

\begin{figure}[t!]
\center
\includegraphics[width=8.5 cm]{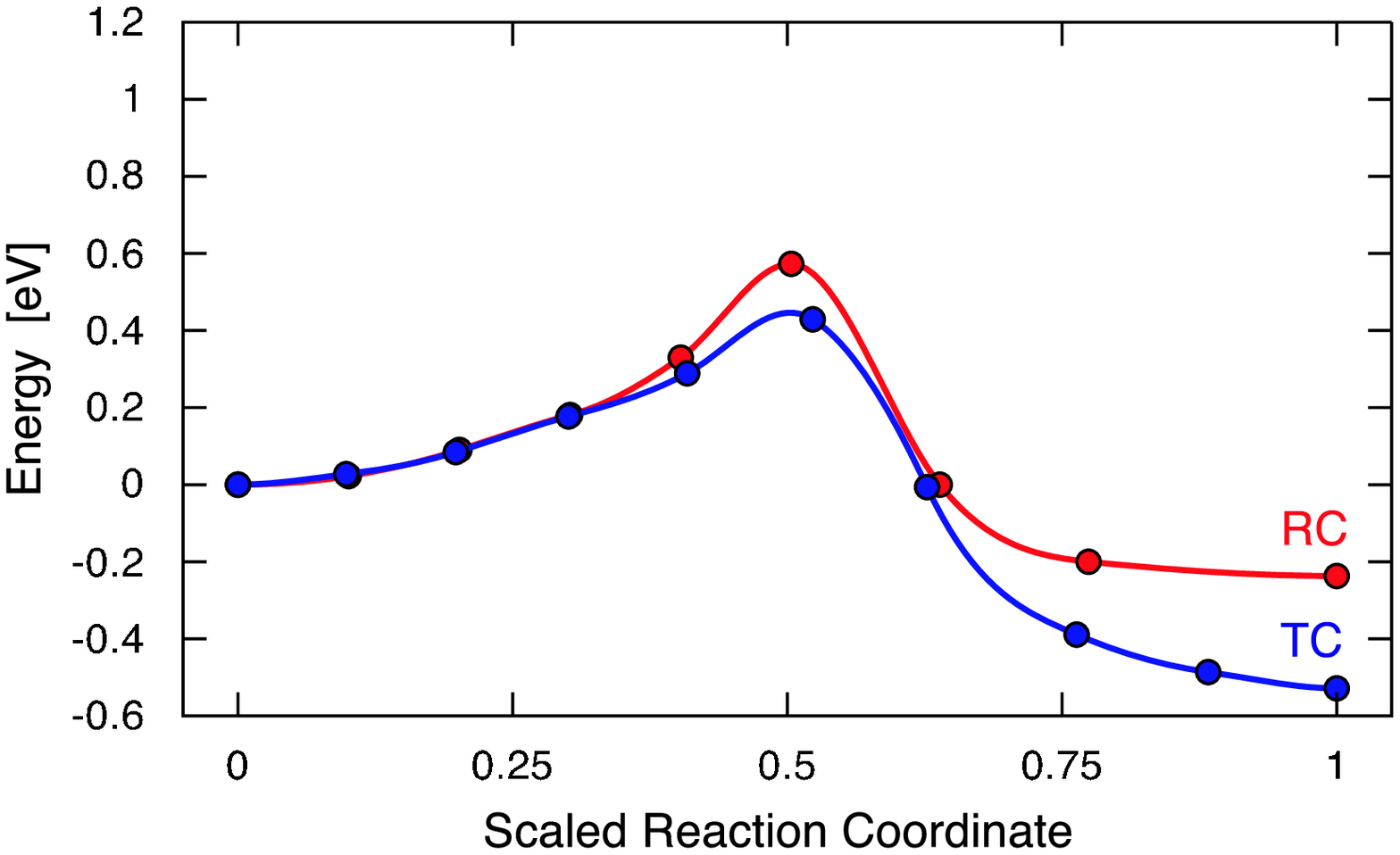} 
\includegraphics[width=8.5 cm]{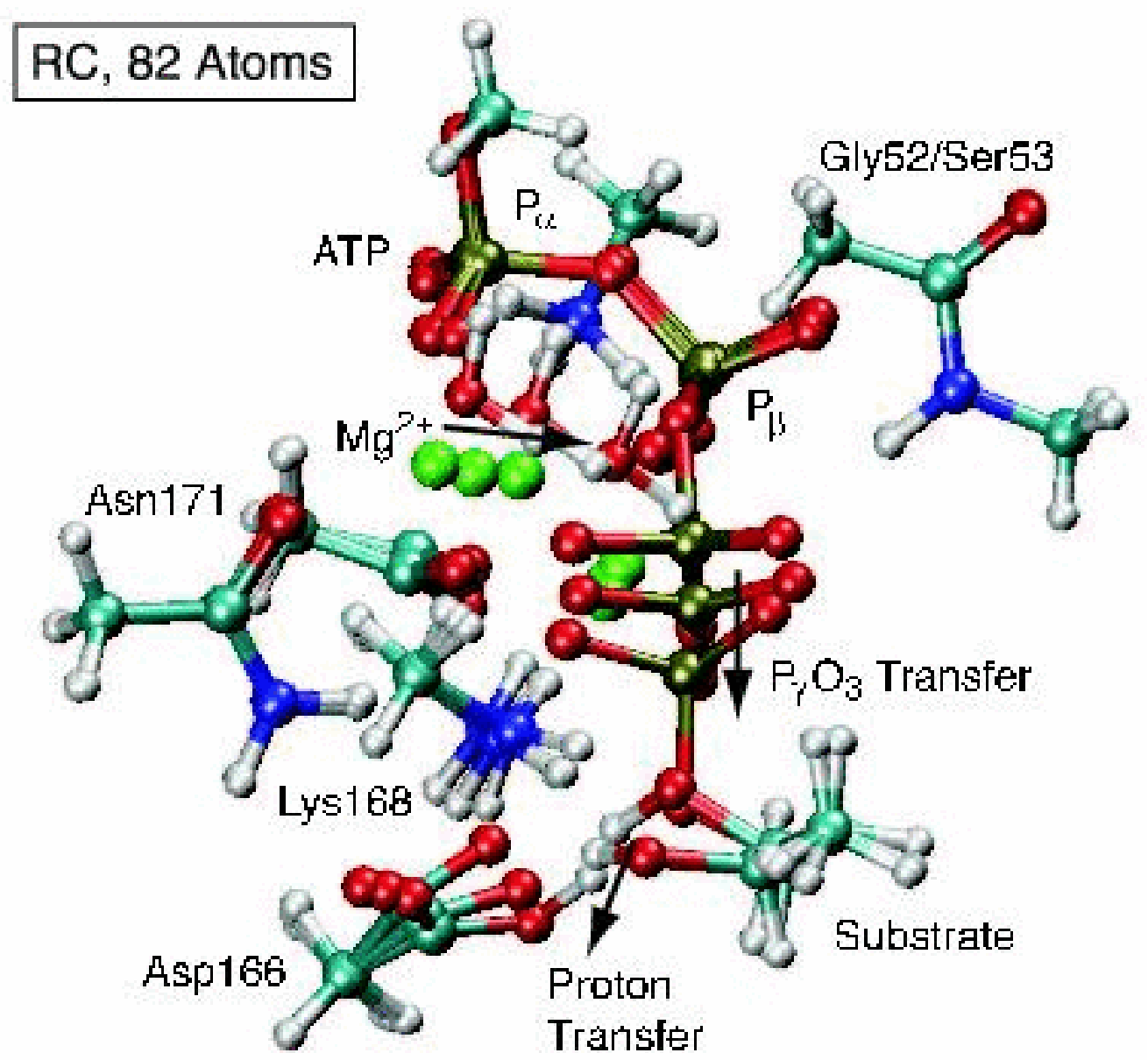} 
\caption{(a) Energy barrier for the 82-atom system in the RC and 
TC structures.  The TC transition
state analog lowers the reaction barrier and favors the product
state. The TC energies are shifted down by 400 meV relative to the
RC energies to facilitate comparison. Each circle represents an image in the
NEB calculation.  (b) Reaction path for phosphate transfer in RS PKA with 82
atoms.  The three image sequence is the reactant state, transition
state, and product state. The atoms are color-coded: Red is oxygen,
light blue is carbon, dark blue is nitrogen, silver is hydrogen, green
is magnesium and gold is phosphorus At the transition state the
phosphate group is planar and the substrate proton has not yet
transfered to the Asp166 catalytic base. 
}
\label{pka82_fig} 
\end{figure}

To check this posibility, we duplicated their 82-atom system and
choice of constraints, and show the results of our calculation in
Fig.\ \ref{pka82_fig}, for both RC and TC.  Consistent with their
result, we find a 0.5 eV barrier and an exothermicity of 0.2~eV in the
RC, indicating a robustness of the result to the various truncation
proceedures, basis sets, periodic boundary conditions, and functionals
used the calculations.  Unfortunately, the result is also insensitive
to the starting conformation of the protein, in contrast to the
results on the 244-atom systems.

Table I shows that the motions of the atoms are much larger than in
the 244-atom calculation, and comparision of Figs. \ref{244atom}b and
\ref{pka82_fig}b shows a qualitatively larger and more contorted
reaction dynamics in the smaller system.  Further reflection suggests
that atoms are moving that would not be able to in the complete
protein system, for example the Mg$^{2+}$ ions.  We tested this by
adding two crystallographic water molecules to Mg$_1$, and observing
the reaction to become endothermic by $\sim 0.5$~eV.  Clearly, the
82-atom system is under-constrained because it is not large enough to
provide accurate energetics of the relative reaction rate of the two
systems.

A potentially very useful observation is that the geometries of
reaction in the 82-atom calculation for both conformations share
several similarities with the 244-atom TC calculation.  This suggests
that the underconstrained calculations can offer useful clues as to
which conformations will favor reaction.

An energetic offset of 400~meV has been subtracted from TC for ease of
comparison of the barrier and exothermicity to RC.  Although we do not
place great confidence in this number, it in interesting to note that
the additional interactions included in the 244-atom system stabilize
the reactive conformation of the protein.

\section {DISCUSSION}

Much has been written about the reaction dynamics problem
\cite{warshel03, hanggi90, frauenfelder85, zheng96}, but it is
primarily concerned with the very difficult problem of combining the
vibrational transition with solvent motions.
These ideas lead to something like
\cite{florian03}, which does not make use of the separation of energy
scales which comes from understanding protein dynamics, {\em per se},
and does not provide modularity of the different aspects of the
calculation (quantum mechanical, hydration, and conformational
motions.) By allowing relatively few atoms to move, we increase
the probability that the dynamics of traversing our
zero-temperature pathway will be simple enough to occur quickly. 
The formalism of Agmon and Hopfield can then be used to
relate the vibrational transition to an overall reaction rate.

\begin{figure}[t!]
\center
\includegraphics[width=6.5 cm]{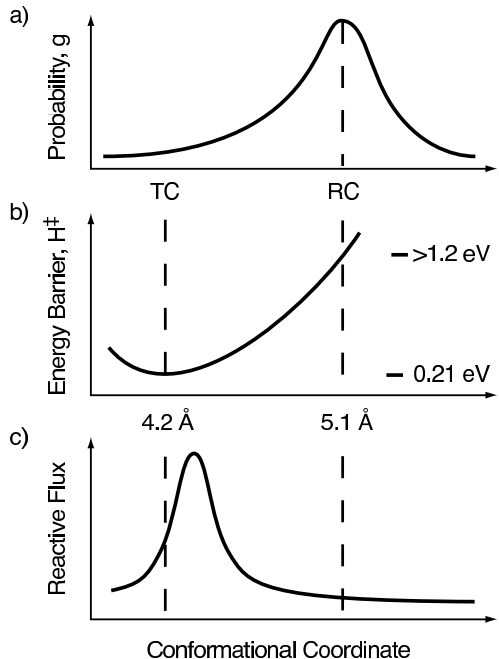} 
\caption{ Schematic separation of reaction rate into conformational and
vibrational (Arrhenius) coordinates.  {\bf (a)}, The distribution of protein
conformations, with RC representing an average structure and TC a 
less-populated member of the distribution.  {\bf (b)} We have calculated
an enthalpy barrier for two members of the ensemble, and interpolated
between in this figure.  {\bf (c)} Reactive flux as a function of conformation,
calculated using the Agmon-Hopfield formalism, described in the text.}
\label{ratefig} 
\end{figure}

The current calculations are evaluated in the context of the
Agmon-Hopfield model in Figure 4a, which shows the 1ATP (RC) structure
at the peak of a distribution of conformations and 1L3R (TC) at the
side.  Figure \ref{ratefig}b shows the barrier computed in the two
calculations --- 60 $k_BT$ for the reactant conformer and 6 $k_BT$
for the transition conformer.  Figure \ref{ratefig}c shows the
reaction flux across the barrier as a function of protein
conformation, simply the product of the probability of a conformation
with the rate of reaction for that conformation $g\left( cc
\right)k\left( cc \right)$, where $k\left( cc\right) =
\exp\left[-H\left(cc\right)/k_BT\right]$. 

This model provides an estimate of the distance-scale over which the
barrier changes can be made by combining the information in Figs.\
\ref{pka82mg_fig} and \ref{ratefig}c, which shows that a 0.9 \AA\
shift in the Mg$^{2+}$ position, combined with several other changes
of a similar size shifts the barrier by 50 $k_BT$.  Thus, a reasonable
flux is confined to a multi-dimensional region only $\sim 0.2$ \AA\
wide!  Allostery is the property by which small molecules or proteins
binding distant from the active site can influence activity by changing
the protein conformation; the present calculation shows that these changes
can be quite sublte.

The exact value of the reaction rate requires knowlege of the conformational
occupancy of the low-barrier conformation, which we have not attempted to
calculate in this work.  There is a need for methods that can explore
conformational changes in proteins.  In this regard, inspection of the
calculation in Fig. 3 can be quite helpful in discovering what
conformations to look for when screening an MD simulation to find the
correct portion of conformation space.

A recent crystal structure of a Y204A mutation of PKA in complex with
a peptide inhibitor supports two aspects of our calculation \cite{yang04}.
First, a mixture of reactants and products can co-exist, and second, very
few residues move in the course of the phosphate transfer.  Yang {\em et al.}
specifically note the absence of motion in the Mg$^{++}$ ions, coordinated
water, K168, and S53 \cite{yang04}.

\section {CONCLUSIONS}

Two important lessons can be learned from this work.  First, approximately
200 atoms need to be included to obtain the correct electronic structure, 
and also maintain enough constraints on the boundary conditions to
meaningfully reproduce the effect of protein conformation on reaction
rate.  Second 
is the existance of the zero-temperature pathway in this particular
case.  In proteins where it is not possible to obtain TC crystal structures,
it is essential to find appropriate methods to explore the variety of 
protein conformational motions.  Since protein motions typically occur
on timescales ranging from hundreds of nanoseconds to microseconds,
it is unlikely that simple embedding of the quantum region in a larger 
classical region will resolve this difficulty.

Is {\em ab-initio} quantum chemistry now in a position to answer the biologically
motivated questions posed in the introduction?  It is clear, at least, that the
active site cleft is {\bf not} closed far enough in the RC structure.  On the
other hand, the real power of these techniques will only become evident when
it is possible to thouroughly sample conformational motions of proteins and 
protein complexes.  If studies of protein folding are any guide, this day is
fast-approaching \cite{clementi03}.

\section {ACKNOWLEDGMENTS}

We thank Matt Challacombe, Angel Garc\'{\i}a, John Portman, Art Voter, and
Hans Frauenfelder for valuable discussions and Matt Challacombe
for obtaining the necessary computer time on the QSC supercomputer at
Los Alamos National Lab, which made the large simulations possible.
GH would also like to acknowledge Eric Galburt for suggesting that the
NEB method should be used to look at reaction mechanisms in biological
systems \cite{galburt01}.  
This work was performed with support from the Department of Energy, 
under contract W-7405-ENG-36 and the Laboratory Directed Research and
Development program at Los Alamos National Laboratory.


\newpage
\bibliographystyle{prsty}

\end{document}